\documentstyle[twocolumn,prl,aps,epsfig]{revtex}
\begin{document}
\flushbottom
\draft
\title{Wave atom optics theory of the Collective Atomic-Recoil Laser}
\author{M. G. Moore and P. Meystre}
\address{Optical Sciences Center and Department of Physics\\
University of Arizona, Tucson, Arizona 85721\\
(November 25, 1997)
\\ \medskip}\author{\small\parbox{14.2cm}{\small \hspace*{3mm}
We present a wave atom optics theory of the Collective Atomic-Recoil Laser,
where the atomic center-of-mass motion is treated quantum mechanically.
It extends the previous ray atom optics theory, which treated the 
center-of-mass atomic motion classically, to the realm of ultracold atoms.
For the case of a far off-resonant pump laser we derive an analytical
solution which gives the linear response of the CARL system for both the 
quantum and classical regimes. A linear stability analysis reveals significant 
qualitative differences between these two regimes, which arise from the effects 
of diffraction on the atomic center-of-mass motion.
\\[3pt]PACS numbers: 42.55-f,42.50.Vk,03.75.-b  }}
\address{}\maketitle
\maketitle
\narrowtext
The Collective Atomic-Recoil Laser, or CARL, is the atomic analog
of the Free Electron Laser \cite{Bra90}. It relies on the interplay between
atomic motion in light fields and the dependence of these fields on the
spatial atomic distribution to generate coherent light from an atomic gas
in the absence of population inversion. Similarly to the situation for FEL,
gain is correlated with the appearance of a density grating in the atomic
sample, i.e. bunching. 

The theory of collective atomic-recoil lasers was developed by 
Bonifacio et al \cite{BonSalNar94,SalCanBon95}, who described
the atoms as classical point particles moving in the optical potential
provided by the light fields. A number of their predictions were
experimentally verified by Bigelow et al \cite{HemBigKat96}, using a hot
atomic sample for which this ``ray atom optics'' approach is certainly
justified. Additional experiments by Courtois and coworkers
\cite{CouGryLou93} using cold cesium atoms, and by Lippi et al
\cite{LipBarBar96} using hot sodium atoms measured the recoil-induced
small-signal probe gain, however these experiments did not include
probe feedback, which is an essential part of the CARL system.

The purpose of this letter is to extend the theory of the CARL to the regime
of ``wave atom optics'', where the atomic center-of-mass motion is treated
quantum mechanically. This extension is of importance for experiments using
ultracold atomic samples. \footnote{ What we call the ``ray'' and ``wave''
atom optics regimes are sometimes called the semiclassical and quantum
regimes in the laser cooling literature. We prefer to reserve this
terminology for the standard quantum optics use, where the semiclassical
regime refers to a classical description of light.} We show that in this
regime the small-signal behavior of the CARL is qualitatively vastly
different from its ray optics counterpart. This is a consequence of
atomic diffraction, which counteracts the bunching process, and thus tends
to inhibit the gain.  

A simple way to discuss the CARL is in terms of
pump-probe spectroscopy, with the understanding that in practice quantum
noise, rather than a weak probe, will trigger lasing. CARLs operate by
placing an atomic vapor in the field of a strong pump laser, which in
conjunction with a weak counterpropagating probe laser results in the
appearance of a periodic optical potential (light shift). As a result of
the associated mechanical forces an initially homogeneous sample
acquires a density modulation at the period of the optical potential.
Amplification can therefore be interpreted as stimulated scattering of the
pump beam off the atomic density grating. The amplified probe beam is
then fed back into the atomic sample via a ring cavity. Hence any increase
in the probe strength results in a stronger standing wave, and thus more
bunching and increased scattering of the pump laser. This runaway
amplification is finally reversed by saturation effects.

The wave optics theory of a CARL laser is similar to that of atomic
diffraction by standing waves \cite{BerSho81}, except that the
electromagnetic field must now be treated as a dynamical variable.
It is also similar to the theory of recoil induced resonances 
\cite{GuoBerDub92}, which, however, does not include the crucial effects 
of probe feedback.
We restrict our considerations to an initially monochromatic (zero
temperature) atomic sample,
as this permits to isolate with particular simplicity the impact of
atomic diffraction on the operation of the CARL. A full model including
both the nonlinear regime and the effects of finite
atomic temperatures will be presented in a detailed paper in preparation.
We note that near-zero temperature atomic samples are now quite realistic
in view of the successful demonstration of Bose-Einstein condensation in
low density atomic samples.

We describe the CARL as an ensemble of identical two-level atoms
interacting with a probe laser of wave number $k_1\hat{\bf z}$,
and a counterpropagating pump laser of wave number $k_2\hat{\bf z}$. We
neglect transverse effects and in addition
consider the situation where the lasers are tuned far from the two-level
resonance, so that we may ignore the effects of spontaneous emission. We
also consider densities low enough that collisions can be ignored.

The Ray Atom Optics (RAO) model of the CARL is derived from the
Hamiltonian
\begin{eqnarray}
\hat{H}_{R}&=&\sum_j^N\left[\frac{p^2_j}{2m}
+\frac{\hbar\omega_0}{2}\hat{\sigma}_{3j}
+i\hbar\left( g_1a_1^\star e^{-ik_1z_j}\hat{\sigma}_{-j}
\right.\right.\nonumber\\
&+&\left.\left.
g_2a_2^\star e^{-ik_2z_j}\hat{\sigma}_{-j}-H.c.\right)\right],
\label{HRAO}
\end{eqnarray}
where $g_1$ and $g_2$ are the atom-field electric dipole
coupling constants, given by $g_i=\mu[ck_i/(2\hbar \epsilon_0 V)]^{1/2}$,
$i=1,2$, $\mu$ is the magnitude of the atomic dipole moment, and $V$ is
the quantisation volume. The light fields are treated classically,
and $a_1$ and $a_2$ are the normal variables corresponding to the probe
and pump field modes respectively.  The classical variables
$p_j$ and $z_j$ are the momentum and position of the $j^{th}$ atom.
They obey the canonical equations of motion
$dz_j/dt=\partial \hat{H}_{R}/\partial p_j$ and $dp_j/dt=
-\partial \hat{H}_{R}/\partial z_j$. As usual, $\hat{\sigma}_{3j}$ and
$\hat{\sigma}_{\pm j}$ are Pauli pseudo-spin operators which describe the
internal state of the $j_{th}$ two-level atom. They obey the Heisenberg
equations of motion $d\hat{\bbox \sigma}_j/dt=
(i/\hbar)[\hat{H}_{R},\hat{\bbox \sigma}_j]$.
Due to the fact that quantum operators only appear linearly, taking
expectation values of these equations allows one to replace all quantum 
operators by their c-number counterparts.

In the absence of collisions, the Wave-Atom-Optics (WAO) description of
the CARL is most easily obtained from the single-particle Hamiltonian
\begin{eqnarray}
\hat{H}_{W}&=&\frac{\hat{p}^2}{2m}
+\frac{\hbar\omega_0}{2}\hat{\sigma}_z
+i\hbar\left[g_1a_1^\star\ e^{-ik_1\hat{z}}\hat{\sigma}_-
\right.\nonumber\\
&+&\left.
g_2a_2^\star e^{-ik_2\hat{z}}\hat{\sigma}_--H.c.\right],
\label{HW}
\end{eqnarray}
where $\hat{p}$ is the center-of-mass momentum operator and
$\hat{z}$ the center-of-mass position operator, with $[{\hat z}, {\hat p}]
= i\hbar$. and $\hat{\sigma}_z$ and $\hat{\sigma}_\pm$ are Pauli pseudo-spin
matrices acting on the internal atomic state. The atomic sample of
$N$ identically prepared atoms is then described by an effective
single-particle density operator, or population matrix $\hat{\rho}$ with
Tr${\hat \rho} = N$. It obeys the Schr\"odinger equation
$\dot{\hat{\rho}}=(i/\hbar)[\hat{\rho},\hat{H}_{W}]$.

In both the RAO and the WAO models the field variables obey Maxwell's
wave equation
\begin{equation}
\frac{d}{dt}a_i=-i\omega_i a_i
+g_iN\langle e^{-ik_i \hat{z}}\hat{\sigma}_-\rangle,
\label{dadt}
\end{equation}
where the source term, proportional to $N$, is the polarization of the 
medium, and we have neglected the effects of cavity losses.
In the RAO model the expectation value is interpreted as a
classical average over individual atoms, while in WAO it is the
quantum expectation value $Tr[{\hat \rho}(t)\exp (-ik_i {\hat z})
{\hat \sigma}_-]$.

We evaluate the small-signal response of the CARL by
linearizing its equations of motion about the exact solution
in the absence of probe field, $a_1=0$. We make the approximations
$k_1\approx -k_2 \approx k_0 = \omega_0/c$, $g_1\approx g_2 = g$, and
assume that the pump field remains undepleted, an approximation appropriate
for the linear regime, so that
$a_2=a_2(0)\exp[-i(\omega_2+\Delta\omega) t]$.
Here $a_2(0)$ is a constant taken to be real without loss of generality,
and $\Delta\omega$ is the frequency shift due to the atomic polarization.

We consider specifically the far-off resonant situation
where the upper electronic state of the atoms is adiabatically eliminated,
a procedure equivalent to describing them as an ensemble of
classical Lorentz atoms. (Note that this approximation requires that we
avoid atomic densities large enough for dispersion to shift the pump
frequency into resonance.)  Since spontaneous emission is neglected, our
model describes therefore a purely dispersive medium where the atoms serve 
as a catalyst to transfer photons from the pump to the probe, but no absorption
can occur. 

While any initial momentum distribution
can be considered, a particularly simple set of equations results for the
case of an initially monochromatic beam of atoms. In the atoms' rest
frame, the dimensionless probe field $A_1 = a_1/a_2$ then obeys the
equation of motion
\begin{equation}
\frac{d}{d\tau}A_1=i(\Delta_{21}A_1+\beta B),
\label{dA1dt}
\end{equation}
where
$\tau=4\omega_r t$, $\Delta_{21}=(\omega_2-\omega_1)/4\omega_r$,
$\beta=g^2N/4\omega_r(\omega_0-\omega_2)$, and we have introduced the recoil
frequency $\omega_r=\hbar k_0^2/2m$. We remark that for a fixed detuning, 
changing $\beta$ simply corresponds to varying the atomic density.
We have introduced the atomic bunching parameter $B$, via 
\begin{equation}
B=\langle e^{-2ik_0\hat{z}}\rangle.
\end{equation}
The magnitude of $B$ is thus a measure of the degree of bunching of the
atomic sample.

Equation (\ref{dA1dt}) is valid in both the
RAO and the WAO regimes. It is the equation of motion for $B$, however,
where the difference becomes apparent.
It reads
\begin{equation}
\frac{d^2}{d\tau^2}B=-\eta B +\alpha A_1,
\label{Beq}
\end{equation}
where $\eta=0$ corresponds to the RAO regime, and $\eta=1$ the WAO regime.
Here $\alpha=2g^2[a_2(0)]^2/4\omega_r(\omega_0-\omega_2)$ so that for fixed
detuning, changing $\alpha$ corresponds to varying the pump
intensity.

Having its origin in the kinetic energy part of the Hamiltonian (\ref{HW}),
the term $-\eta B$ in Eq. (\ref{Beq}) gives the effects of atomic diffraction
on the evolution of $B$.
In our scaled variables $d^2B/d\tau^2$ is proportional to
$1/\omega_r^2$ and $\alpha$ is proportional to $1/\omega_r$, hence the
WAO description reduces to the classical result of the RAO model in the
limit $\omega_r \to 0$,
which in the present situation is equivalent to
taking the limit $m \to \infty$ or $\hbar \to 0$,
a clear demonstration of the
correspondence principle.  Everything else being equal,
massive atoms are better modeled by Ray Atom Optics than lighter atoms since
they suffer less diffraction.

The form of Eq. (\ref{Beq}) shows that the diffraction term $-\eta B$ can
be interpreted as a restoring ``force'' which acts on the bunching.
Unlike the classical regime, where $B$ behaves like a {\em free particle}
driven by the probe field $A_1$, quantum mechanically it behaves like a
{\em simple harmonic oscillator} of frequency $4\omega_r$(in
original time units), and subject to that same driving force. In the
linear regime, the bunching parameter $B$ is assumed to be a small
perturbation around its initial value of zero, and the optical potential
resulting from a nonzero $A_1$ tends to increase it. But this mechanism is
opposed by diffraction in the WAO regime.

Equations (\ref{dA1dt}) and (\ref{Beq}), form
a set of coupled-mode equations for the probe field and the atomic bunching.
In order to precisely compute the probe gain, one would ideally want the
full solution to (\ref{dA1dt}) and (\ref{Beq}), but in this work
we choose to focus on the eigenvalue spectrum, and in particular, the 
regime of linear instability. The time-dependent probe gain will be included 
in a more detailed paper.  The spectrum for the system is
obtained by solving the cubic equation
\begin{equation}
\lambda^3-i\Delta_{21}\lambda^2+\eta\lambda
-i(\alpha\beta+\eta\Delta_{21})=0.
\label{cubic}
\end{equation}
With the substitution $\lambda=i\tilde{\lambda}$, Eq.
(\ref{cubic}) reduces to a cubic equation for $\tilde{\lambda}$ with real
coefficients. The set of eigenvalues $\{\lambda\}$ can fall into one of
two categories: (I) all eigenvalues are imaginary, and (II) one eigenvalue
is imaginary, and the other two are complex, with equal and opposite real parts.
Case (I) is stable, and no linear growth of the probe field can occur,
but case (II) implies that precisely one eigenvalue has a positive real
part. In that case, the system grows exponentially from its initial state. 
This growth is characterized by a rate $\Gamma$
given by the largest real part of the eigenvalue spectrum.
From Eq. (\ref{cubic}) it becomes immediately apparent that the eigenvalue 
spectrum depends only on two control parameters: $\Delta_{21}$ and the product
$\alpha\beta=2g^4N[a_2(0)]^2/16\omega_r^2(\omega_0-\omega_2)^2$.

From these considerations, it is possible to introduce a threshold
condition, defined as the point of transition between cases I and II. It
is given by
\begin{eqnarray}
\left(\frac{\alpha\beta}{2}\right)^2+\alpha\beta\frac{\Delta_{21}}{3}
\left[\eta-\left(\frac{\Delta_{21}}{3}\right)^2\right]\nonumber\\
-\eta\left[\frac{1}{27}+\frac{2}{3}\left(\frac{\Delta_{21}}{3}\right)^2
-2\left(\frac{\Delta_{21}}{3}\right)^4\right] > 0.
\label{thresh}
\end{eqnarray}
For the RAO model ($\eta=0$) this reduces to
$\alpha\beta>4\Delta_{21}^3/27$
and above threshold the growth rate is given by
\begin{equation}
\Gamma_R=\frac{\sqrt{3}}{2}\left(\frac{\alpha\beta}{4}\right)^{1/3}
\left|(1+\sqrt{d})^{2/3}-(1-\sqrt{d})^{2/3}\right|,
\label{RAOsoln}
\end{equation}
where $d=1-4\Delta_{21}^3/27\alpha\beta$, and the subscript $R$ stands for
Ray Atom Optics.
Figure 1(a) plots $\Gamma_R$ versus $\Delta_{21}$
for various values of the product $\alpha\beta$.
This figure agrees with the numerical results obtained by Bonifacio et al,
see e.g. Figs. 3(b) and 4(b) of Ref. \cite{SalCanBon95}.
\centerline{\psfig{figure=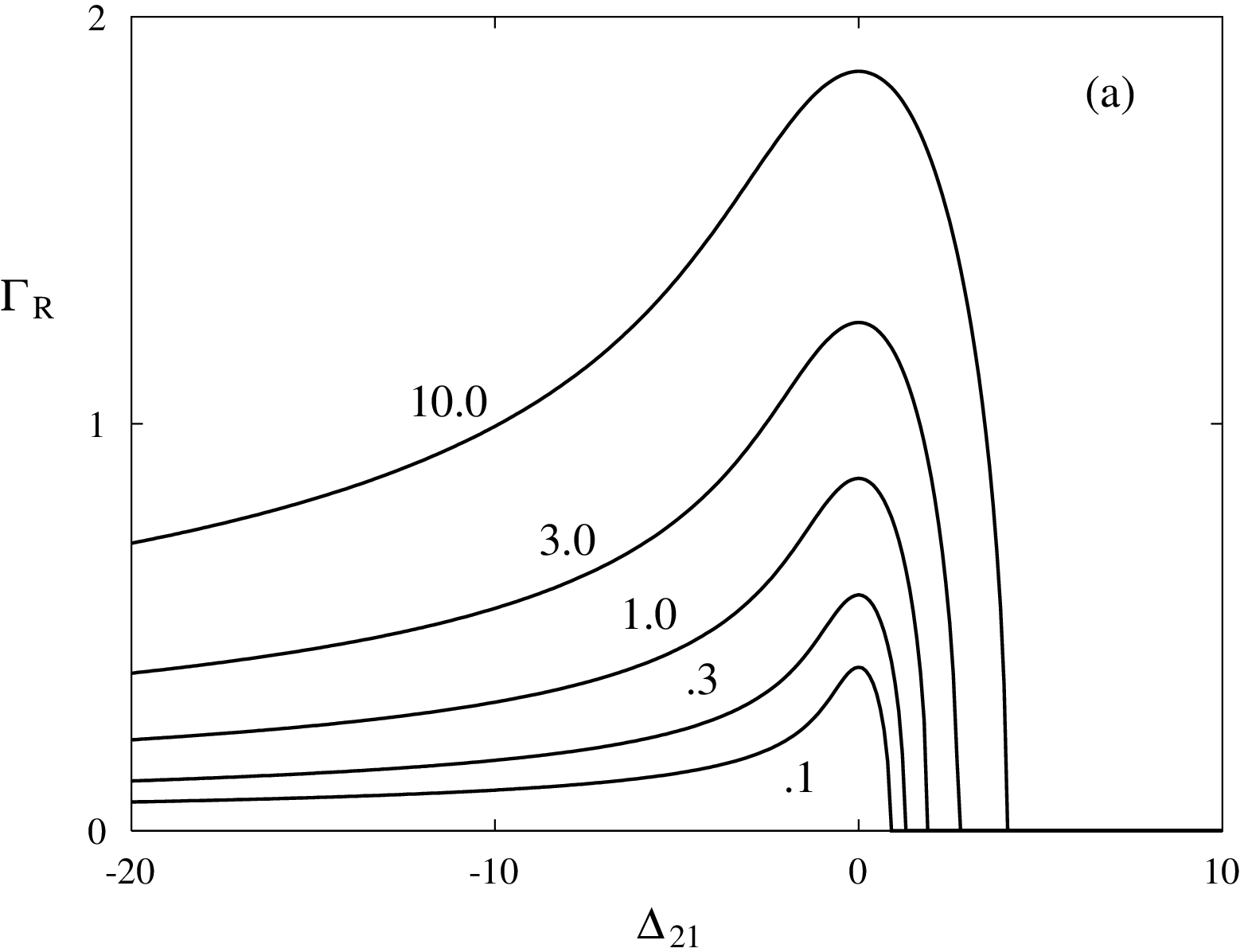,width=8.6cm,clip=}}
\centerline{\psfig{figure=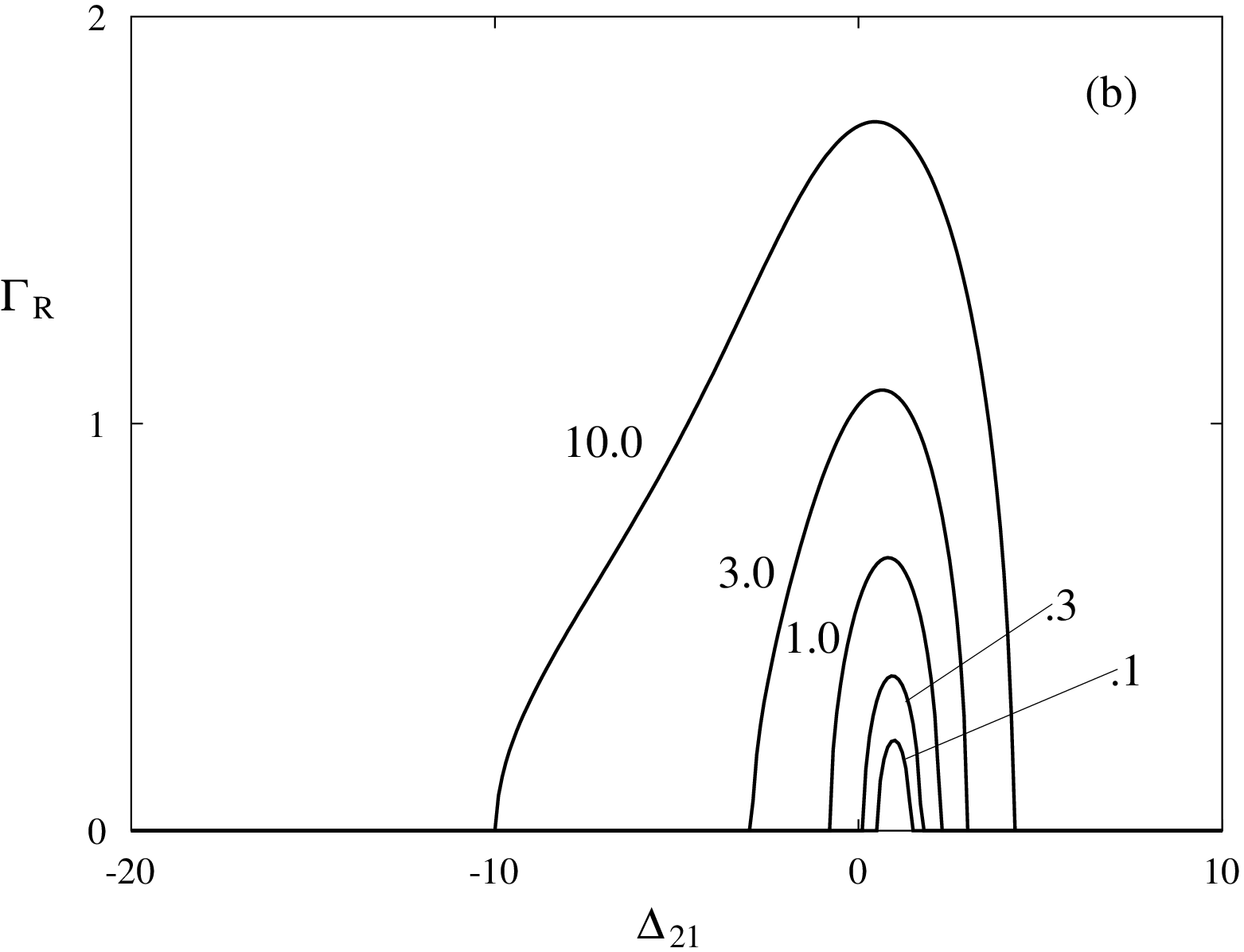,width=8.6cm,clip=}}
\begin{figure}
\caption{Comparison of the linear growth rate versus pump-probe 
detuning between the RAO and WAO regimes.  For each curve, 
the value of the product $\alpha\beta$ is given. Fig. 1a shows $\Gamma_R$, 
while Fig. 1b shows $\Gamma_W$.}
\end{figure}

Figure 1(a) should be compared with Fig. 1(b), which plots the analytical
WAO growth rate $\Gamma_W$ versus $\Delta_{21}$ for the same parameters. 
(The explicit form of $\Gamma_W$ is cumbersome, and we do not
reproduce it here.) Figures 1(a) and (b) show many qualitative differences
between the predictions of the RAO and WAO models. The most striking is
the appearance of a second threshold: for values of the pump-probe
detuning $\Delta_{21}$ below a critical value, atomic diffraction inhibits
the creation of a density modulation in the sample, and the system
exhibits no gain. In addition, we observe a shift in the position of maximum
gain away from $\Delta_{21}=0$ as the atomic density decreases.
In the limit of weak pump intensities and/or low atomic densities
the behavior shown in Figs. 1(a) and (b) can be understood
quite simply.  The atomic center-of-mass dispersion curve tells us that 
the absorption of a pump photon and the emission
of a probe photon by an atom initially at rest creates an energy
defect of $4\omega_r$ due to atomic recoil.  This defect can be compensated
by a detuning between the pump and probe, thus the fact that $\Gamma_W$
is non-zero for only a small range around $\omega_2-\omega_1=4\omega_r$ 
is simply an expression of energy-momentum conservation.  
In contrast, the RAO model
has its maximum value at $\omega_2-\omega_1=0$, consistent with
the RAO limit $\omega_r \to 0$, i.e. it assumes that
the center-of-mass dispersion curve is essentially flat.

As expected from our earlier discussion, and with all other parameters
fixed, the differences between the RAO and WAO descriptions become less
pronounced as the atomic mass increases. This is illustrated in Fig. 2,
which plots $\Gamma_R$ and $\Gamma_W$ versus $\Delta_{21}$ for
$\alpha\beta=5(m/m_0)^2$ for three different values of the atomic mass.
This corresponds to varying the mass while holding the pump intensity and
atomic density constant.
The classical and quantum models do indeed converge as $m$ increases, as
predicted by the correspondence principle. Note however that the mass does
not appear explicitly in the coupled-mode equations (\ref{dA1dt})
and (\ref{Beq}). This means that whatever the
atomic mass may be, there will be some set of parameters where the
ray atom optics description of the CARL ceases to be
valid.
\centerline{\psfig{figure=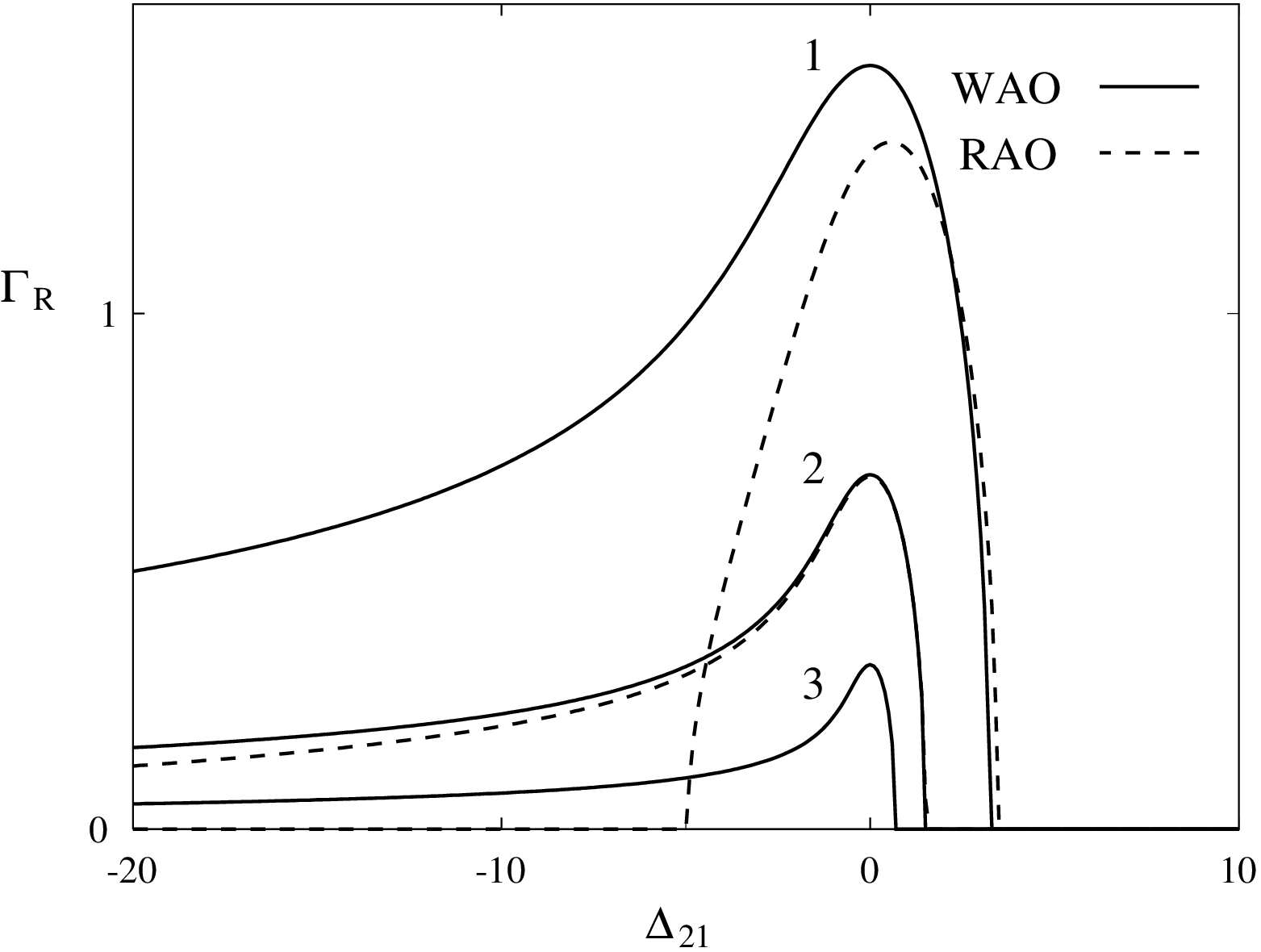,width=8.6cm,clip=}}
\begin{figure}
\caption{The linear growth rate as a function of pump-probe
detuning, showing the effect of increasing atomic mass.
The solid line shows $\Gamma_W$, while the dashed line shows $\Gamma_R$. 
With $\alpha\beta=5(m/m_0)^2$, the curves labeled $1$ show the results for 
$m=m_0$.  The curves labeled $2$ and $3$ show the results for
$m=10m_0$ and $100m_0$ respectively. All curves are in units corresponding 
to $m=m_0$.}
\end{figure}

One special case of some interest is the presence of gain at
$\Delta_{21}=0$. From Figs. 1(a) and (b) we can compare $\Gamma_R$ and
$\Gamma_W$ for this case. While the RAO  model
predicts gain for all values of $\alpha\beta$, as can also be inferred 
from Eq. (\ref{thresh}), WAO predicts a threshold, a direct consequence 
of matter-wave diffraction, given by
$\alpha\beta > 2/3\sqrt{3}$.   

In summary, a comparison between a ray atom optics and a wave atom optics
description of the CARL illustrates the fundamental role of matter wave
diffraction in the qualitative behavior of the system. In particular, it
leads to the appearance of new thresholds resulting
from the competition between matter-waves diffraction and the spatial
modulation of the atomic density by the optical potential.

\acknowledgements
We have benefited from discussions with R. Bonifacio and L. De Salvo, who
brought the CARL system to our attention. This work is supported in part 
by the U.S. Office of Naval Research Contract No. 14-91-J1205, by the 
National Science Foundation Grant PHY95-07639, by the U.S. Army Research 
Office and by the Joint Services Optics Program.

\end{document}